\documentclass[aps,prl,reprint,showpacs,twocolumn,superscriptaddress,floatfix]{revtex4-1}

\usepackage{graphicx}
\usepackage{amssymb}

\newcommand*{\HamT}{{\cal{H}}_T}

\newcommand*{\Heff}{\mbox{\boldmath$H$}}
\newcommand*{\SMM}{\mbox{\boldmath$S$}}

\newcommand*{\svec}{\mbox{\boldmath$\sigma$}}

\newcommand*{\sy}{\sigma_y}

\newcommand*{\jvec}{\mbox{\boldmath$j$}}

\begin{document}

\title{Spin-polarized Shapiro steps and spin-precession-assisted multiple Andreev reflection} 

\author{C. Holmqvist}

\affiliation{Fachbereich Physik, Universit\"at Konstanz, D-78457 Konstanz, Germany}

\author{M. Fogelstr\"om}

\affiliation{Department of Microtechnology and Nanoscience - MC2, Chalmers University of Technology,
SE-412 96 G\"oteborg, Sweden}

\author{W. Belzig}

\affiliation{Fachbereich Physik, Universit\"at Konstanz, D-78457 Konstanz, Germany}



\date{\today}

\begin{abstract}
We investigate the charge and spin transport of a voltage-biased superconducting point contact coupled to a nanomagnet.
The magnetization of the nanomagnet is assumed to precess with the Larmor frequency, $\omega_L$, due to ferromagnetic resonance.
The interplay between the ac Josephson current and the magnetization dynamics leads to spin-polarized Shapiro steps at voltages $|V|=\hbar\omega_L/2en$ for $n=1,2,...$ and the subharmonic steps with $n>1$ are a consequence of multiple Andreev reflection (MAR). Moreover, the spin-precession-assisted MAR generates quasiparticle scattering amplitudes that, due to interference, lead to current-voltage characteristics of the dc charge and spin currents with subharmonic gap structures displaying an even-odd effect.
\end{abstract}


\maketitle


{\it Introduction}.
Control of the electron's spin degree of freedom has led to a number of spintronics applications. \cite{tserkovnyak2005} These applications often rely on using the spontaneously broken spin-rotation symmetry of a ferromagnet to act as a reference for the spin of itinerant electrons.
In contrast to ferromagnets, 
the 
electron-electron interaction in a BCS superconductor leads to a spinless order parameter. 
While these two states of matter usually are incompatible in bulk materials, they can be combined in nanoscale junctions to create new phenomena. \cite{eschrig2011,bergeret2005,buzdin2005} The interplay between superconductivity and ferromagnetism in such hybrid superconducting junctions allows for probing the spin degree of freedom as well as for coherently controlling spin-polarized currents. \cite{fogelstrom2000,zhao2004,andersson2002,zhao2007,cuevas2001,bobkova2006,keizer2006,senapati2011,khaire2010,huertas-hernando2002,shomali2011} In addition, superconducting junctions coupled with magnetization dynamics have been
studied \cite{volkov2009,zhu2003,michelsen2008}.
The lowest-energy magnetization excitation can be accessed by the application of an external magnetic field, which starts a precession of the magnetization around the direction of the field \cite{bell2008,morten2008,skadsem2011,richard2012}.
This ferromagnetic resonance (FMR) mode can also be achieved by a coupling between the magnetization and the Josephson effect: the ac Josephson current generates an oscillating magnetic field that resonantly excites the magnetization precession \cite{thirion2003,petkovic2009,barnes2011}.

In Ref.~\cite{petkovic2009}, 
the coupling between a spatially dependent order parameter and the magnetization dynamics was studied in a tunnel junction whose width, $W$, was larger than the superconducting coherence length, $\xi_0$, and this coupling was detected as a rectification of the ac Josephson charge current. However, in the regime $W<\xi_0$, which is the case for a superconducting quantum point contact (SQPC) \cite{muller1992,scheer1997}, the spin-rotation symmetry of the superconducting order parameter may be broken locally and one may anticipate that spin-dependent modifications of the superconducting correlations will influence the transport properties \cite{teber2010,holmqvist2011}.
In this Letter, we study the coupling of the ac Josephson effect and magnetization dynamics of an SQPC containing a nanomagnet,  e.g. a magnetic impurity \cite{thirion2003} or a single-molecule magnet \cite{bogani2008,heersche2006,jo2006,kasumov2005,kahle2012}.
We show that the interplay between the Josephson and magnetization oscillations create Shapiro-like resonances that lead to a rectified spin current. In addition, we find a rich subgap structure of the current-voltage characteristics of the dc charge and spin currents displaying features related to the emission or absorption of energy quanta corresponding to the precession frequency.
However, the features associated with an odd number of Andreev reflections are suppressed due to interference.

The transport properties of nonmagnetic SQPCs can be understood in terms of Andreev reflection. 
An applied bias voltage leads to the occurrence of multiple MAR \cite{octavio1983} and the ac Josephson effect, which is characterized by the Josephson frequency, $\omega_J=2eV/\hbar$.
The ac current of SQPCs also includes higher harmonics of the Josephson frequency, i.e. $j(t)=\sum_n j_n e^{in\omega_Jt}$ (setting  $\hbar=1$) \cite{bratus1995,averin1995,cuevas1996}.
In addition, MAR processes produce a subharmonic gap structure of the dc current that can be seen in the current-voltage characteristics as steps at voltages $V=2\Delta/en$, $n=1,2,\dots$ 
The ac Josephson effect can be probed by the application of microwave radiation as their interplay leads to Shapiro steps and a rich subgap structure due to photon-assisted MAR \cite{cuevas2002,uzawa2005,chauvin2006}.
In the case of SQPCs with magnetic interfaces, the 
transport properties are modified due to spin-dependent effects \cite{fogelstrom2000,andersson2002,cuevas2001}. E.g., a magnetic interface may produce different transmission probabilities for the spin-up and spin-down bands, ${\cal D}_\uparrow$ and ${\cal D}_\downarrow$, which may lead to the occurrence of spin currents 
\cite{bobkova2006,zhao2007}.

Here, we consider a point contact consisting of two superconducting leads coupled over a nanomagnet.
The left (L) and right (R) leads are assumed to consist of s-wave superconductors in the clean limit \cite{cuevas1996}.
We further assume that the magnetization of the nanomagnet can be treated classically within the macrospin model and hence be described by single spin vector, $\SMM$ \cite{tserkovnyak2005}.
Under FMR conditions, external dc and rf fields are applied to create an effective magnetic field, $\Heff$, in which the spin precesses with the  Larmor frequency $\omega_L=\gamma | \Heff |$, where $\gamma$ is the gyromagnetic ratio.
This effective field also includes any effects of the resonant coupling between the ac Josephson current and the magnetization dynamics.
The nanomagnet's spin, whose dynamics can be described by the Landau-Lifshitz-Gilbert equation \cite{LLG}, generates a time-dependent exchange field that is felt by the quasiparticles tunneling across the junction and can be incorporated into a phenomenological tunnel Hamiltonian that allows for 
spin flip scattering $\downarrow\rightarrow\uparrow$ ($\uparrow\rightarrow\downarrow$) which is accompanied by absorption (emission) of an energy quantum $\omega_L$ by the quasiparticle
(see Ref.~\cite{holmqvist2011} for details).

The transport characteristics depend on
the time-dependent superconducting phase difference across the contact, $\varphi(t)=\varphi_0+\omega_J t$, where $\varphi_0$ is the initial phase difference, in addition to the Larmor frequency.
The spin $\SMM$ traces out an in-plane angle described by $\chi(t)=\chi_0 + \omega_L t$ in analogy with $\varphi(t)$.
The scattering resulting from the MARs combined with the precession-induced spin flips transfers quasiparticles into sidebands with energies $\varepsilon_n^m\equiv \varepsilon+n\omega_J+m\omega_L$, where $n=0,\pm1,\pm2,\dots$ and $m=0,\pm1$.
The charge ($\mu=0$) and spin ($\mu=x,y,z$-component) currents in lead $\alpha=L,R$ can then be written in terms of all harmonics as
\begin{eqnarray}
j^\mu_\alpha(t) = \int\frac{d\varepsilon}{2\pi} \sum_{n,m} e^{-i(n\varphi_0+m\chi_0)-i(n\omega_J+m\omega_L)t} (j^\mu_\alpha)^m_{n}\,. \label{eq:jmu}
\end{eqnarray}

{\it Model}.
The currents are obtained using nonequilibrium Green's function techniques and the calculation is based on the Hamiltonian approach \cite{cuevas1996} in which the leads are described by Keldysh Green's functions in the quasiclassical approximation \cite{eliashberg1972} and the interface containing the nanomagnet is treated as a strong impurity \cite{serene1983}. 
The effects of $\varphi(t)$
are accounted for in the standard way by a transformation of the unperturbed lead Green's functions, $\check{g}_\alpha$, where "$\check{\;\;}$" denotes a matrix in Keldysh space, that is expressed as $\hat{{\cal U}}_\alpha=\exp[i\varphi_\alpha(t) \hat{\tau}_3/2]$ in Nambu space ("$\hat{\;\;}$").
The voltage-biased lead Green's functions, $\check{g}_\alpha(t,t')$, are then expressed in terms of the equilibrium ones, $\check{g}_\alpha(t-t')$, as $\hat{g}^{X}_\alpha(t,t')=\hat{{\cal U}}_\alpha^\dagger(t)\hat{g}^{X}_\alpha(t-t')\hat{{\cal U}}_\alpha(t')$ where $X$ refers to one of the retarded ($R$), advanced ($A$) or Keldysh ($K$) components.
The equilibrium Green's functions have a simple Fourier transformation given by $\check{g}_\alpha(t-t')=\int (d\varepsilon/2\pi) \, e^{-i\varepsilon(t-t')}\check{g}_\alpha(\varepsilon) $. Since the leads consist of s-wave superconductors, the retarded component of $\check{g}_\alpha(\varepsilon)$ is given by
$\hat{g}^R_\alpha = \hat{\tau}_3 g^R(\varepsilon) + \hat{\tau}_1 f^R(\varepsilon) i \sigma_y $,
where $g^R(\varepsilon)=-\pi \varepsilon/\Omega$, $f^R(\varepsilon)=-\pi\Delta /\Omega$, $\Omega=\sqrt{|\Delta|^2-(\varepsilon+i 0)^2 }$. 
The advanced component is $\hat{g}^A_\alpha(\varepsilon)=\hat{\tau}_3 [ \hat{g}^R_\alpha(\varepsilon) ]^\dagger \hat{\tau}_3$ while the Keldysh Green's function is given by $\hat{g}^K_\alpha(\varepsilon)=[\hat{g}^R_\alpha(\varepsilon) - \hat{g}^A_\alpha(\varepsilon)] \tanh (\varepsilon/2T)$.

\begin{figure}[t] 
\includegraphics[width=0.99\columnwidth,angle=0]{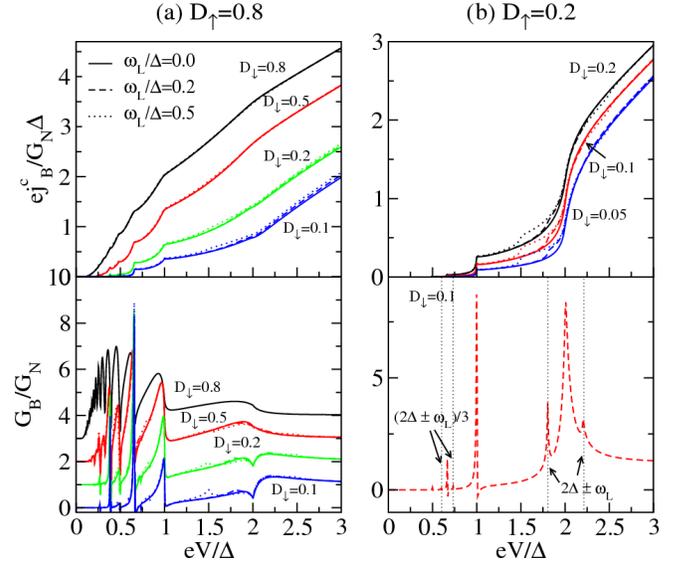}
\caption{\label{fig: background current}
(Color online) Charge background current and conductance, $G_B=\partial j^c_B/\partial V$, for a $\pi$ junction with (a) ${\cal D}_\uparrow=0.8$ and (b)  ${\cal D}_\uparrow=0.2$. The current is normalized by the normal conductance given by the two spin channels, $G_N=[e^2/h] [{\cal D}_\uparrow + {\cal D}_\downarrow]$.
The conductance curves in panel (a) have been offset with 3 (${\cal D}_\downarrow=0.8$), 2 (${\cal D}_\downarrow=0.5)$ and 1 (${\cal D}_\downarrow=0.2$).
In panel (b), the conductance is plotted for  ${\cal D}_\uparrow=0.2$, ${\cal D}_\downarrow=0.1$ and $\omega_L/\Delta=0.2$.
In all plots, $\vartheta=\pi/8$ and the temperature is $T=0$.}
\end{figure}

\begin{figure}[t] 
\includegraphics[width=0.9\columnwidth,angle=0]{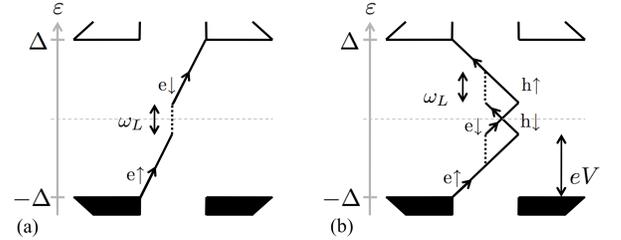}
\caption{\label{fig: Larmor process}
(a) First-order tunneling process in which a spin-up electron-like quasiparticle, e$\uparrow$, absorbs a precession quantum $\omega_L$ and undergoes a spin flip. (b) Second-order process which leads to destructive interference of the combined Andreev reflection amplitudes of the hole-like quasiparticles h$\uparrow$ and h$\downarrow$.}
\end{figure}



The magnetic interface is characterized by a tunnel Hamiltonian, $\HamT= \sum_{k\sigma; k^\prime \sigma^\prime} c^\dagger_{L,k\sigma} v_{L,k\sigma;R,k^\prime\sigma^\prime} c_{R,k^\prime \sigma^\prime} +h.c.$, in which the hopping amplitudes, $\check{v}_{LR}(t)=\check{v}_{RL}(t)=\check{v}(t)=\check{v}^\dagger(t)$, depend on the instantaneous direction of the spin, $\hat{\SMM} (t)$.
Within the quasiclassical approximation,
the Fermi surface limits of the hopping amplitudes can be expressed as $\hat{v}_{k; k^\prime }=(v_0 +v_S(\hat{\SMM} (t) \cdot \svec))\delta(k-k^\prime)$. 
We can define the spin-up (spin-down) transmission probability as ${\cal D}_{\uparrow(\downarrow)}=4v_{\uparrow(\downarrow)}^2/[1+v_{\uparrow(\downarrow)}^2]^2$, where $v_{\uparrow(\downarrow)}=v_0 \pm v_S \cos\vartheta$ and $\vartheta$ is the angle between $\SMM$ and $\Heff$.

The hopping elements' influence on the lead Green's functions is captured by the t-matrix equation
$\check{t}_\alpha(t,t')=\check{\Gamma}_\alpha(t,t')+ \int dt_1 \int dt_2 \check{\Gamma}_\alpha(t,t_1)\check{g}_\alpha(t_1,t_2)\check{t}_\alpha(t_2,t')$, where $\check{\Gamma}_{L/R}(t,t')=\check{v}_{LR/RL}(t)\check{g}_{R/L}(t,t')\check{v}_{RL/LR}(t')$.
The time dependence of the superconducting phase of the Green's functions can be transferred to the hopping elements by the transformation $\hat{{\cal U}}_\alpha(t)$ according to $\hat{{\cal U}}_{L/R}(t)\hat{v}_{LR/RL}(t)\hat{{\cal U}}^\dagger_{R/L}(t)=\exp[\pm i \varphi(t) \hat{\tau}_3/2]\hat{v}(t)$.
The $t$-matrix equation then becomes an algebraic equation in energy space
and 
\begin{eqnarray}
\check{t}_\alpha(t,t') &=& \sum_{n,m} e^{-i(n\varphi_0+m\chi_0)+i(n\omega_J+m\omega_L)t'} [\check{t}_\alpha(t-t')]^m_n,
\nonumber\\
\,[\check{t}_\alpha(t)]^m_n  &=&  \int \frac{d\epsilon}{2\pi}e^{-i\varepsilon t}[\check{t}_\alpha(\epsilon^m_n)].
\end{eqnarray}
By only considering currents on the left side of the junction (thereby dropping the index $\alpha=L$), one can write 
the Fourier components as $\check{t}_{n}^{m}=[\check{t}(\varepsilon^m_n)]$ 
and the current components of Eq.~(\ref{eq:jmu}) as
\begin{equation}
(j^\mu)^m_{n} =  \int \frac{d\varepsilon}{4\pi}{\rm Tr}\{\hat{\kappa}^\mu[  \check{t}_{n}^{m} \check{g}_0^0-\check{g}_n^m  \check{t}_{n}^{m}  ]^< \},
\end{equation}
where $\check{g}^m_n=\check{g}(\varepsilon_n^m)$, $\hat\kappa^0=e\hat\tau_3$, $\hat{\kappa}^i={\rm diag} (\sigma_i, \sigma_y\sigma_i\sigma_y)/2$ for $i=x,y,z$. The $t$-matrix components can be found from the equation
$\check{t}^m_{n} = \check{\Gamma}^m_{n}  + \sum_l \{ \check{A}^{m,l}_{ n,n} \check{t}^l_{ n} +  \check{B}^{m,l}_{n,n+1} \check{t}^l_{n+1} +   \check{B}^{m,l}_{n,n-1} \check{t}^l_{n-1}   \}$
where the retarded components are given by
\begin{widetext}
\begin{eqnarray}
&&[\hat{\Gamma}^R]^{m}_{n} = \sum_j \left(\begin{array}{cc}  \nu^{m-j} \nu^{j} [g^R]^{j}_{-\frac{1}{2}}  \,  \delta_{n,0} & e^{-i\varphi_0} \nu^{m-j} i\sy (\nu^{j})^\dagger  [f^R]^{j}_{\frac{1}{2}} \,   \delta_{n,1}  \\
e^{i\varphi_0} (\nu^{m-j})^\dagger i\sy  \nu^{j} [f^R]^{j}_{-\frac{1}{2}}  \, \delta_{n,-1}  &  -(\nu^{m-j})^\dagger (\nu^{j})^\dagger  [g^R]^{j}_{\frac{1}{2}}  \, \delta_{n,0}  \end{array}\right)  ,  
\\ 
&&[\hat{A}^R]_{n,n}^{m,l} =  \sum_{j=-2}^2 \left(\begin{array}{cc}  \nu^{m-j}  \nu^{j-l} [g^R]^j_{n-\frac{1}{2}} [g^R]^l_{n} &  \nu^{m-j}  \nu^{j-l} i\sy [g^R]^j_{n-\frac{1}{2}} [f^R]^l_{n} \\
-(\nu^{m-j})^\dagger (\nu^{j-l})^\dagger i \sy [g^R]^j_{n+\frac{1}{2}} [f^R]^l_{n} &
(\nu^{m-j})^\dagger (\nu^{j-l})^\dagger  [g^R]^j_{n+\frac{1}{2}} [g^R]^l_{n} \end{array}\right)  , 
\\
&&[\hat{B}^R]_{n,n+1}^{m,l} = \sum_{j=-2}^2  e^{i\varphi_0} (\nu^{m-j})^\dagger  i \sy   \nu^{j-l} [f^R]^j_{n+\frac{1}{2}}  \left(\begin{array}{cc} 0 & 0 \\
( g^R )^l_{n+1} & [f^R]^l_{n+1} i\sy \end{array}\right) 
\\ 
{\rm and} \quad &&[\hat{B}^R]_{n,n-1}^{m,l} =  \sum_{j=-2}^2 e^{-i\varphi_0}  \nu^{m-j}i\sy (\nu^{j-l})^\dagger [f^R]^j_{n-\frac{1}{2}}  \left(\begin{array}{cc} i\sy [f^R]^l_{n-1} & -[g^R]^l_{n-1} \\
0 & 0 \end{array}\right),
\end{eqnarray}
\end{widetext}
where we have defined $\nu^0=v_0+v_S\sigma_z\cos\vartheta$, $\nu^{\pm 1}=\frac{v_S}{2}[\sigma_x\pm i\sigma_y]\sin\vartheta $ and $\nu^j=0$ if $|j|\ge 2$. The advanced and Keldysh matrices are similarly defined.

\begin{figure}[b] 
\includegraphics[width=0.98\columnwidth,angle=0]{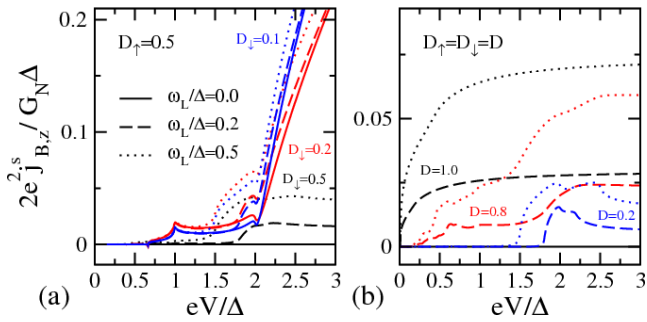}
\caption{\label{fig: spin background current}
(Color online) The $z$ component of the dc background spin current. 
In all plots, $\vartheta=\pi/8$ and $T=0$.}
\end{figure}

{\it Results}.~--
The dc charge and spin currents of Eq.~(\ref{eq:jmu}) consist primarily of contributions from the 
$(j^\mu)^0_0$
components and can be described as background currents 
in the spirit of Ref.~\cite{cuevas2002}.
There is also the possibility of Shapiro-like resonances between the Josephson and the Larmor frequencies at voltages $V_n^m=-(m/n)\omega_L/2e$, $m \neq 0$.
The Larmor precession produces a spin structure of the $t$ matrices in which the index $m=1(-1)$ corresponds to a spin-raising (spin-lowering) matrix. This spin structure 
causes the Shapiro resonances to give contributions to the spin-polarized current only and thus result in a Shapiro spin current given by $\jvec^s_{{\rm Shapiro}}=\sum_{n,m=\pm1}e^{-i(n\varphi_0+m\chi_0) }(\jvec^s)^m_n \delta(V-V_n^m)$. This behavior is in contrast to that of a weak link coupled to a nanomagnet whose magnetization performs a precessional motion. In that case, the resulting time-dependent magnetic field produces Shapiro-type resonances that give contributions to the 
charge current \cite{cai2010}.

Figure \ref{fig: background current} shows the current-voltage characteristics as well as the conductance for the charge background current, $j^c_B$.
As can be seen, the most prominent feature 
is the subharmonic gap structure appearing at voltages $eV=(2\Delta\pm\omega_L)/n$, where $n=1,3,...$
This structure can clearly be seen in the conductance where it is displayed as distinct satellite peaks for junctions in the tunnel limit.
The subharmonic gap structure is similar to that of the current-voltage characteristics generated by photon-assisted MAR which shows features due to the absorption or emission of photons \cite{cuevas2002}.
However, in the case of MAR processes influenced by Larmor precession, a quasiparticle can only absorb or emit one quantum of $\omega_L$ since these processes are accompanied by a spin flip (see e.g.~Fig.~\ref{fig: Larmor process} for the absorption processes with $n=1$ and $2$). As a result, there is only one satellite peak on each side of the feature corresponding to a tunnel process of order $n$.
In addition, only the processes for which $n$ is an odd number display side peaks.
One can show that this suppression is due to destructive interference of the Andreev reflection amplitudes corresponding to the side peaks of processes with $n=2,4,\dots$
Defining $\delta V^\pm=[eV-2\Delta\pm \omega_L]/2\Delta$, the height of the current steps at $\delta V^\pm  \ll 1$ can be approximated in the tunnel limit by
$j^c_\pm \approx \Theta\left(\delta V^\pm \right)   e\Delta \frac{2}{\pi}v_s^2\sin^2\vartheta  I\left( [eV\pm\omega_L]/2\Delta \right)$, where $\Theta$ is the Heaviside step function, $I(a)=[ 2a E( \sqrt{1-1/ a^2})   - K ( \sqrt{ 1-1/a^2})/a ]$ and $K$ and $E$ are the complete elliptic integrals of the first and second kinds.
In the limit $\omega_L\ll2\Delta$, one obtains $I\left( [eV\pm\omega_L]/2\Delta \right) \approx  \frac{\pi}{2}  \left[ 1+ \frac{3}{2} \delta V^\pm  \right]$ and the heights of the current steps at $V=(2\Delta\pm\omega_L)/e$ are thus $\propto v_s^2\sin^2\vartheta$.

The $z$ component of the background spin current, $\jvec^s_B$, is shown in Fig.~\ref{fig: spin background current} (the in-plane components are zero).
For transmission probabilities ${\cal D}_\uparrow \neq {\cal D}_\downarrow$ and $\omega_L=0$, our results reproduce those of Ref.~\cite{zhao2007} for zero spin mixing.
In the case of nonzero Larmor frequency, the spin current is finite even for ${\cal D}_\uparrow = {\cal D}_\downarrow$, as can be seen in panel (b).
In the tunnel limit, the spin current can be divided into a spin-filter current and a spin-pump current, $\jvec^s_B=\jvec^s_{\rm filter}+\jvec^s_{\rm pump}$. The $z$ component of the spin-filter current is given by
$j^s_{{\rm filter},z}= (\Delta/2\pi) ({\cal D}_\uparrow-{\cal D}_\downarrow) I(eV/2\Delta)$
for $eV\geq 2\Delta$ \cite{zhao2007}. 
The spin-pump current can correspondingly be approximated by
$j^s_{{\rm pump},z,\pm}=\pm j^c_\pm/2e$.
For $eV>2\Delta+\omega_L$ and $\omega_L\ll\Delta$, the total spin-pump current is given by $j^s_{{\rm pump},z}= (3/4)v_s^2 \sin^2\vartheta \, \omega_L$ if $\omega_L\ll\Delta$ and is hence $\propto\omega_L$ and does not depend on the bias voltage.

\begin{figure}[t] 
\includegraphics[width=1.0\columnwidth,angle=0]{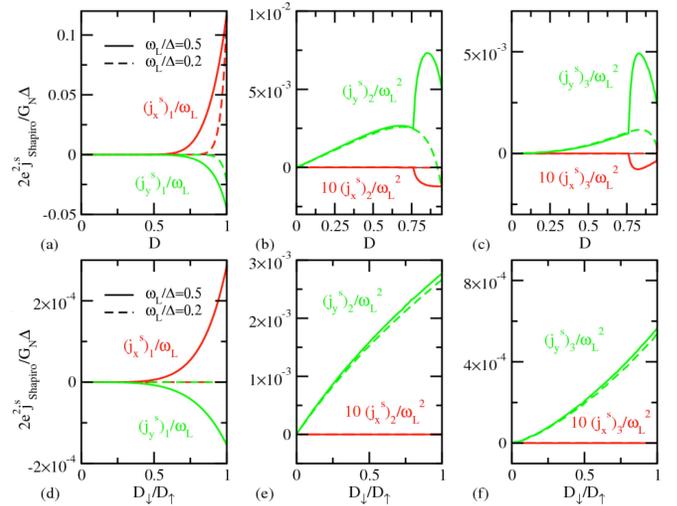}
\caption{\label{fig: shapiro}
(Color online) The Shapiro-resonance contributions (a) $(\jvec^s)_1$, (b) $(\jvec^s)_2$, and (c) $(\jvec^s)_3$ for ${\cal D}_\uparrow = {\cal D}_\downarrow = {\cal D}$. Panels (d)-(f) show the Shapiro-resonance contributions (d) $(\jvec^s)_1$, (e) $(\jvec^s)_2$, and (f) $(\jvec^s)_3$ for ${\cal D}_\uparrow = 0.5$.
In all plots, $\vartheta=\pi/8$ and $T=0$.}
\end{figure}

Now, we turn to the Shapiro-like resonances in the spin current. Since the Larmor precession requires $m$ to take the values $m=\pm 1$ for the Shapiro resonance condition,
$\jvec^s_{\rm Shapiro}$ is
spin polarized in the $xy$ plane (see Ref.~\cite{holmqvist2011} for a detailed analysis of the $t$ matrices' spin structure).
Defining $(\jvec^s)_{n} \delta(V-V_n) \equiv  \sum_{m=\pm1} [(\jvec^s)^{m}_{n}+(\jvec^s)^{m}_{- n}] \delta(V-V^m_n)$ where $V_n=\pm \omega_L/2en$, one can write the $n$th component as
$(\jvec^s_{{\rm Shapiro}})_n = \hat{R}^\dagger (\jvec^s)_{n} \hat{R} \, \delta(V-V_n)$,
where $n\geq 1$ and the transformation $\hat{R}=\exp[\frac{i}{2} (-n\varphi_0\pm \chi_0) \hat{\Heff} \cdot \svec ]$ rotates the vector $(\jvec^s)_{n}$ around the direction of the external magnetic field, $\hat{\Heff}$, through an angle 
$- n\varphi_0 + {\rm sign}(V_n) \chi_0$.
The existence of a dc current that is spin polarized in the $xy$ plane implies that the rotational symmetry around the $z$ axis is broken and that the magnitude of the Shapiro currents depend on the initial angle of nanomagnet's magnetization direction. This behavior is analogous to the dependence on the initial superconducting phase difference which also is present in the behavior of microwave-irradiated Josephson junctions \cite{cuevas2002}.
The Shapiro contributions strongly depend on the transmission probability as can be seen in Fig.~\ref{fig: shapiro}.

The $\omega_L$ values in Fig.~\ref{fig: background current} are chosen for clarity but may be smaller in practice. Consider an SQCP consisting of Nb, whose superconducting gap is $\Delta\sim 1$ meV, containing a nanomagnet that under FMR conditions reaches precession frequencies of $\omega_L\sim 10$ GHz which is well below the critical magnetic field, $H_c$, of Nb \cite{bell2008}. Then, typically $\omega_L/\Delta=0.01$  and the temperature is restricted to $T<100$ mK.
Increasing $\Heff$ closer to the critical field value decreases $\Delta$ and hence increases the ratio $\omega_L/\Delta$ which allows for a better resolution of the subgap structures of $j^c_B$ and $\jvec^s_B$.
Alternatively, using the ac Josephson current to resonantly excite $\SMM(t)$, $\omega_L$ values corresponding to an effective magnetic field $|\Heff|>H_c $ can be achieved \cite{petkovic2009,barnes2011,thirion2003}. Detection of the spin-polarized Shapiro currents would then be a measurement of the FMR frequency of the nanomagnet.

{\it Conclusions}.~--
In conclusion,
we have calculated the dc charge and spin currents through a voltage-biased superconducting point contact coupled to
the spin of a nanomagnet that under ferromagnetic resonance conditions precesses with the Larmor frequency $\omega_L$.
We have shown that coherent multiple Andreev reflection in the presence of the Larmor dynamics leads to
a subharmonic gap structure of the dc charge and spin currents with features at $eV=(2\Delta\pm\omega_L)/n$ originating from the absorption or emission of a precession quantum. However, destructive interference suppresses these features for even values of $n$.
In addition, the Larmor dynamics combined with broken spin-rotation symmetry generates Shapiro-like resonances resulting in an additional spin current polarized in the $xy$ plane.

{\it Acknowledgments}.~--
We wish to thank J.C. Cuevas for useful discussions. C.H. and W.B. were supported by the DFG and SFB 767. M.F. acknowledges support from the Swedish Research Council (VR).

\end{document}